\documentclass[twocolumn,trackchanges]{aastex631}
\usepackage{xcolor}
\usepackage{amsmath}

\shorttitle{Multicomponent Activity Cycles using Hilbert-Huang Analysis}
\shortauthors{Velloso et al.}

\graphicspath{{./}{figures/}}

\begin{document}

\title{Multicomponent Activity Cycles using Hilbert-Huang Analysis}

\author[0000-0002-9668-5547]{E.~N.~Velloso}
\affiliation{Universidade Federal do Rio Grande do Norte (UFRN), 59078-970, Natal, RN, Brazil}

\author[0000-0003-1963-636X]{F. Anthony}
\affiliation{Universidade Federal do Rio Grande do Norte (UFRN), 59078-970, Natal, RN, Brazil}

\author[0000-0001-7804-2145]{J.-D.~do~Nascimento,~Jr.}
\affiliation{Universidade Federal do Rio Grande do Norte (UFRN),  59078-970, Natal, RN, Brazil}
\affiliation{Center for Astrophysics $|$ Harvard $\&$ Smithsonian, 60 Garden Street, Cambridge, MA 02138, USA}

\author[0000-0002-7146-4916]{L.~F.~Q. Silveira}
\affiliation{Universidade Federal do Rio Grande do Norte (UFRN), 59078-970, Natal, RN, Brazil}

\author{J.~Hall}
\affiliation{Lowell Observatory, 1400 West Mars Hill Road, Flagstaff, AZ 86001, USA}

\author[0000-0001-7032-8480]{S. H. Saar}
\affiliation{Center for Astrophysics $|$ Harvard $\&$ Smithsonian, 60 Garden Street, Cambridge, MA 02138, USA}

\begin{abstract}
The temporal analysis of stellar activity evolution is usually dominated by a complex trade-off between model complexity and interpretability, often by neglecting the non-stationary nature of the process.
Recent studies appear to indicate that the presence of multiple coexisting cycles in a single star is more common than previously thought.
The correct identification of physically meaningful cyclic components in spectroscopic time series is therefore a crucial task, which cannot overlook local behaviors.
Here we propose a decomposition technique which adaptively recovers amplitude- and frequency-varying components.
We present our results for the solar activity as measured both by the sunspot number and the $K$-line emission index, and we consistently recover the Schwabe and Gleissberg cycles as well as the Gnevyshev-Ohl pattern probably related to the Hale cycle.
We also recover the known 8-year cycle for 61 Cygni A, in addition to evidence of a three-cycles long pattern reminiscent of the Gnevyshev-Ohl rule. This is particularly interesting as we cannot discard the possibility of a relationship between the measured field polarity reversals and this Hale-like periodicity.
\end{abstract}


\section{Introduction} \label{sec:intro}

The number of spots that emerge on the solar surface is known to vary in a regular 11-year cycle \citep{Schwabe1844}.
Polarimetric observations of the Sun have revealed the relationship between these structures and the solar magnetic field \citep{Hale1908}, with the interesting observation that the magnetic polarity of sunspots switches sign every consecutive cycle. This polarity reversal means that a complete magnetic cycle spans 22-years, twice the length of the Schwabe (amplitude) cycle.

A stellar monitoring survey was conducted at Mount Wilson Observatory for over 30 years and led the way towards detecting similar starspot cycles in the chromospheric activity of solar-type stars, as measured by the \ion{Ca}{2} H+K flux \citep{Wilson1978}.
For the many main-sequence stars in which such cyclic behavior was observed, \cite{Brandenburg1998} identified two relationships between cycle length, rotation period and mean activity level, known as the active and inactive branches.
However, many stars since then have been found to present a coexistence of long and short activity cycles, some of these falling on both branches simultaneously, and some between them \citep[e.g.,][]{BoroSaikia2018}.

For a few stars, the large-scale magnetic field has been reconstructed using Zeeman Doppler imaging (ZDI) spanning several years, allowing the detection of field topology variations.
On $\xi$~Boo A, $\epsilon$~Eri and HN Peg, such variations were found to be rapid and localized in time \citep{Morgenthaler2012, Jeffers2014, BoroSaikia2015}.
Although $\tau$~Boo was originally thought to reverse its polarity yearly, significantly longer than its 120-day $S$-index cycle, it was found later that this discrepancy  had arisen from poor time sampling.
\cite{Jeffers2018} discovered $\tau$~Boo's 240-day magnetic cycle, the fastest ever observed, which coincided with two complete consecutive chromospheric cycles.
61 Cyg A was the first cool star beyond the Sun where a magnetic cycle was detected in phase with its chromospheric cycle \citep{BoroSaikia2016}; similar results were recently also confirmed in HD~75332 \citep{Brown2021} and 18 Sco. 

In order to better trace the magnetic evolution of solar-type stars, the Mount Wilson dataset has been recently revisited on several occasions, either by extending the data using multiple other surveys while sticking to traditional periodogram methods \citep[e.g.,][]{BoroSaikia2018}, or by introducing new methods to take into account the non-sinusoidal and locally harmonic nature of the cycles \citep[e.g.,][]{Olspert2018}.
Indeed, as is well known for the Sun, stellar activity cycles are not perfectly harmonic, or even strictly periodic; variations in both cycle length and amplitude are common and can make it challenging to interpret the results of methods in the Fourier domain.

As shown by \cite{Olah2016}, time-frequency analysis can be used to reveal multiple and changing cycles in the Mount Wilson survey dataset.
In this letter, we propose to extend this idea by using fully adaptive time-frequency methods with the specific purpose of disentangling multiple physically meaningful components from the data.

The Hilbert-Huang Transform (HHT), based on the Empirical Mode Decomposition \citep[EMD;][]{Huang1998}, is a data-driven technique that addresses this exact issue.
The goal of Hilbert-Huang analysis is to decompose a signal into a finite sum of amplitude-modulated frequency-modulated (AM-FM) oscillations plus some monotonic trend. Such oscillations, known as Intrinsic Mode Functions (IMFs), present well-defined instantaneous values of amplitude and frequency. These can be derived from their analytic signal using the Hilbert transform.

This method of obtaining a time-frequency representation heavily contrasts with other traditional methods from the literature. Linear methods based on inner products with a pre-determined family of signals, such as the Short-Term Fourier Transform (STFT) or the Wavelet Transform, have their resolution limited by the Heisenberg Uncertainty Principle. On the other hand, the HHT is able to precisely track single values of instantaneous frequencies for each component, while also being agnostic to the specific signal shape.

Many improvements to the original EMD algorithm have been proposed to address different issues arising from the original formulation. For a more in-depth review of the HHT and EMD, see \cite{HuangBook1} and \cite{HuangBook2}. In this letter, we use a combination of the Complete Ensemble EMD with Adaptive Noise \citep[CEEMDAN;][]{Colominas2014} and the post-processing steps proposed by \cite{Wu2009}. The instantaneous frequency is then calculated on the resulting modes using the Normalized Hilbert Transform \citep[NHT;][]{Huang2009}.

Hilbert-Huang analysis has been able to extract and characterize periodicities in the solar cycles from many activity indicators, such as the sunspot number \citep{Barnhart2011a, Gao2017}, the 10.7 cm radio flux, and helioseismic frequency shifts \citep{Kolotkov2015}.

The remaining parts of this letter are organized as follows:
in Section \ref{sec:obs} we define our activity time series datasets both for the Sun and for 61 Cyg A; next, Section \ref{sec:methods} details our data-driven multicomponent analysis method as well as the preprocessing steps needed; Section \ref{sec:results} presents the results of our experiments, and we conclude our discussion in Section \ref{sec:conc}. 

\section{Observations} \label{sec:obs}

For the Sun, the first observational dataset used in our analysis as a proxy for magnetic activity is the daily total sunspot number series obtained from the World Data Center for Solar Index and Long-term Solar Observations (WDC-SILSO) at the Royal Observatory of Belgium (\url{http://www.sidc.be/silso/datafiles}). The dataset spans January 1818 to the present. The second dataset is the monthly-averaged \ion{Ca}{2} K emission index composite; derived from the Kodaikanal, Sacramento Peak, and SOLIS/ISS observations \citep{Bertello2016,Bertello2017,Egeland2017}.

The $S$-index time series data for K dwarf 61 Cyg A is composed of the long-term surveys from the NSO Mount Wilson project \citep[MW;][]{Duncan1991,Baliunas1995} and the solar and stellar activity program from the Lowell observatory \citep[SSS;][]{Hall2007,Lockwood2007}. We also used data published by \cite{BoroSaikia2016} using the NARVAL high-resolution spectropolarimeter  and the data presented in the radial velocity catalog from the California Planet Search \citep[CPS;][]{Howard2010,Rosenthal2021}. We additionally determined the $S$-index from the  ESPaDOnS and NARVAL spectropolarimeter data from 2015 onwards following the methodology described in \cite{Marsden2014}. These measurements span more than 50 years of observations, extending from mid-1967 through early 2018.

\section{Hilbert-Huang Analysis}\label{sec:methods}

The goal of our method is to find a finite set of oscillatory components that add up to the original data, while also describing each individual component by its instantaneous values of amplitude and frequency. To this end, our pipeline consists of a preprocessing step to correct sampling issues, followed by the actual decomposition of the signal, and a post-processing step to ensure the components meet certain conditions needed for the Hilbert analysis to, in the final step, extract their time-frequency representations.

\subsection{Preprocessing}

Although the EMD is a fully data-driven algorithm (meaning that it makes no assumptions a priori on specific signal models), it still performs poorly if the input signal is unevenly sampled \citep[see e.g.,][]{Barnhart2011b}.
Therefore, we follow a similar preprocessing approach to the one used by \cite{Olah2009}, in which the data is first downsampled to uniform bins, and later upsampled with a cubic smoothing spline.

We bin each time series at 90-day intervals using a triangular window to compute the averages and to estimate the variances. The exact value of the binsize has no strong impact on the resulting signal, as long as it's within a reasonable distance from the periods of interest; for stars with wildly different rotation rates, this value should vary accordingly.

The binned data is then interpolated at 10-day intervals by a regularized cubic spline with weights inversely proportional to the bin variances.
The resulting smoothed series preserves the local time-frequency content of the original data to a high extent, as demonstrated by \cite{Olah2009}.
Since there are no significant gaps in the solar  observations, the amount of added information to fill in missing data is negligible. Even for the 61 Cyg A data, which has a few minor gaps, the added information is limited.

\subsection{Decomposition}

The EMD consists of an iterative algorithm in which each mode is successively extracted through a sifting process.
This sifting iteratively updates the current mode by subtracting an estimate of the local mean, corresponding to the average of the upper and lower envelopes, which are taken as the spline interpolations of the local maxima and minima.
Convergence is assessed through a stopping criterion that checks if the resulting IMFs have the following properties:
\begin{enumerate}
    \item the number of local extrema and the number of zero crossings are equal (or differ by one);
    \item the local mean is (close to) zero (almost) everywhere.
\end{enumerate}

In other words, all local maxima must be positive, all local minima must be negative, and their envelopes must be symmetrical. After enough IMFs have been subtracted from the original signal $x(t)$ for the remaining data to be completely monotonic, we can write the resulting decomposition as the finite sum:

\begin{align}
    x(t) = \sum_{k=1}^K C_k(t) + r(t),
\end{align}
where each $C_k(t)$ is one IMF and $r(t)$ is the monotonic trend.

The local nature of EMD may result in mode-mixing, i.e., similar oscillations being separated in different modes or vice-versa.
The improved algorithm used in this work is the Complete Ensemble EMD with Adaptive Noise \citep[CEEMDAN;][]{Colominas2014}.
Its main idea is to apply the EMD to an ensemble of degraded copies of the original signal, each arising from a different realization of additive white Gaussian noise.
The EMD then behaves more closely to a filter bank and the resulting modes are more regular; the decomposition of spurious modes is also avoided.

Specifically, let $\mathcal{E}_k(\cdot)$ be an operator that returns the $k$-th IMF via EMD, then define $\mathcal{M}(\cdot)$ as the local mean operator:

\begin{align}
    \mathcal{M}(x)=x-\mathcal{E}_1(x)
\end{align}

Also, let $w^{(i)}$ be the $i$-th white noise realization ($i=1,\dots,I$).
Then, in each CEEMDAN iteration, we calculate the ensemble average of the local mean:

\begin{align}
    q_k = \dfrac{1}{I}\sum_{i=1}^I\mathcal{M}(q_{k-1} + \beta_{k-1}\mathcal{E}_k(w^{(i)}))
\end{align}
with $q_0=x$ and where $\beta_k$ are coefficients controlling the relative energy of the introduced noise. 
The $k$-th pseudo-IMF $D_k$ is then found by the difference between consecutive averages:

\begin{align}
    D_k = q_{k-1} - q_k
\end{align}

\subsection{Post-processing}

In the above analysis, we referred to the output of the CEEMDAN algorithm as ``pseudo-IMFs''.
This is due to the fact that a sum of two different IMFs does not in general preserve the necessary properties that define an IMF; in particular, the ensemble averages computed during CEEMDAN cause the resulting $D_k$ to lose such features.

Since these properties are important guarantees for the assumptions that each component is a single AM-FM oscillation with well-defined amplitudes and frequencies, a post-processing step is needed to transform the pseudo-IMFs into proper IMFs $C_k$.

In this letter, we formalize the generic steps proposed by \cite{Wu2009} in the following post-processing algorithm:

\begin{align}
    C_k &= \mathcal{E}_1(D_k + r_{k-1}) \\
    r_k &= r_{k-1} + D_k - C_k = \mathcal{M}(D_k + r_{k-1})
\end{align}
with $r_0=0$. Since consecutive modes present oscillations on similar timescales, this algorithm extracts a proper IMF from each consecutive pair of pseudo-IMFs, resulting in the same number $K$ of components. The final monotonic trend is given by $r(t) = r_K(t) + q_K(t)$, the sum of the final step's trends.

\subsection{The marginal Hilbert spectrum}

The main purpose of our analysis is not only to separate each individual component of the time-series signal, but also to identify their underlying periodicities.
Since the extracted IMFs are not constrained to harmonic tones, but are, by construction, single-mode signals, the instantaneous frequency is a better measure for their time-varying periodicities.

We start by assuming each IMF can be written as an AM-FM oscillation with an arbitrary, slow-varying amplitude and phase functions $a_k(t)$ and $\phi_k(t)$:

\begin{align}\label{eq:imf}
    C_k(t) = a_k(t)\cos(2\pi\phi_k(t))
\end{align}

The demodulation of the AM and FM subcomponents can be done with stable results by using the Local Mean Decomposition \citep[LMD;][]{Smith2005}.
This method is similar in nature to the EMD, but it decomposes a signal into a sum of Product Functions (PFs):

\begin{align}\label{eq:pf}
    x(t) = \sum_{j=1}^{J}A_j(t)F_j(t) + m(t)
\end{align}
where $A_j$ and $F_j$ correspond to the AM and FM factors of each PF and $m$ is the local mean residue.
Due to the nature of the stopping criteria for the EMD, equation \ref{eq:imf} is only true as an approximation, while equation \ref{eq:pf} holds whenever the LMD converges; however, the LMD can diverge when processing arbitrarily complex signals, which is why it is only applied here to the already simplified IMFs.
Other examples of this joint EMD-LMD technique can be found in \cite{Yue2020}.

A single PF can thus very closely approximate each IMF, and we can write:

\begin{align}
    C_k(t) = A_k(t)F_k(t) + \mu_k(t)
\end{align}
where $\mu_k(t)$ is very close to zero everywhere, and can locally account for some edge effects of the EMD.
We therefore associate $a_k=A_k$ and $\cos(2\pi\phi_k)=F_k$ and derive the time-varying instantaneous frequency $f_k=\phi_k^\prime$ by Hilbert-transforming the FM component \citep{Huang2009}.
The Hilbert spectra of each mode can hence be defined as the distribution of the instantaneous frequencies weighted by the instantaneous squared amplitudes.

\section{Results and Discussion} \label{sec:results}

\subsection{Solar Cycles}

\begin{figure*}[ht!]
\plotone{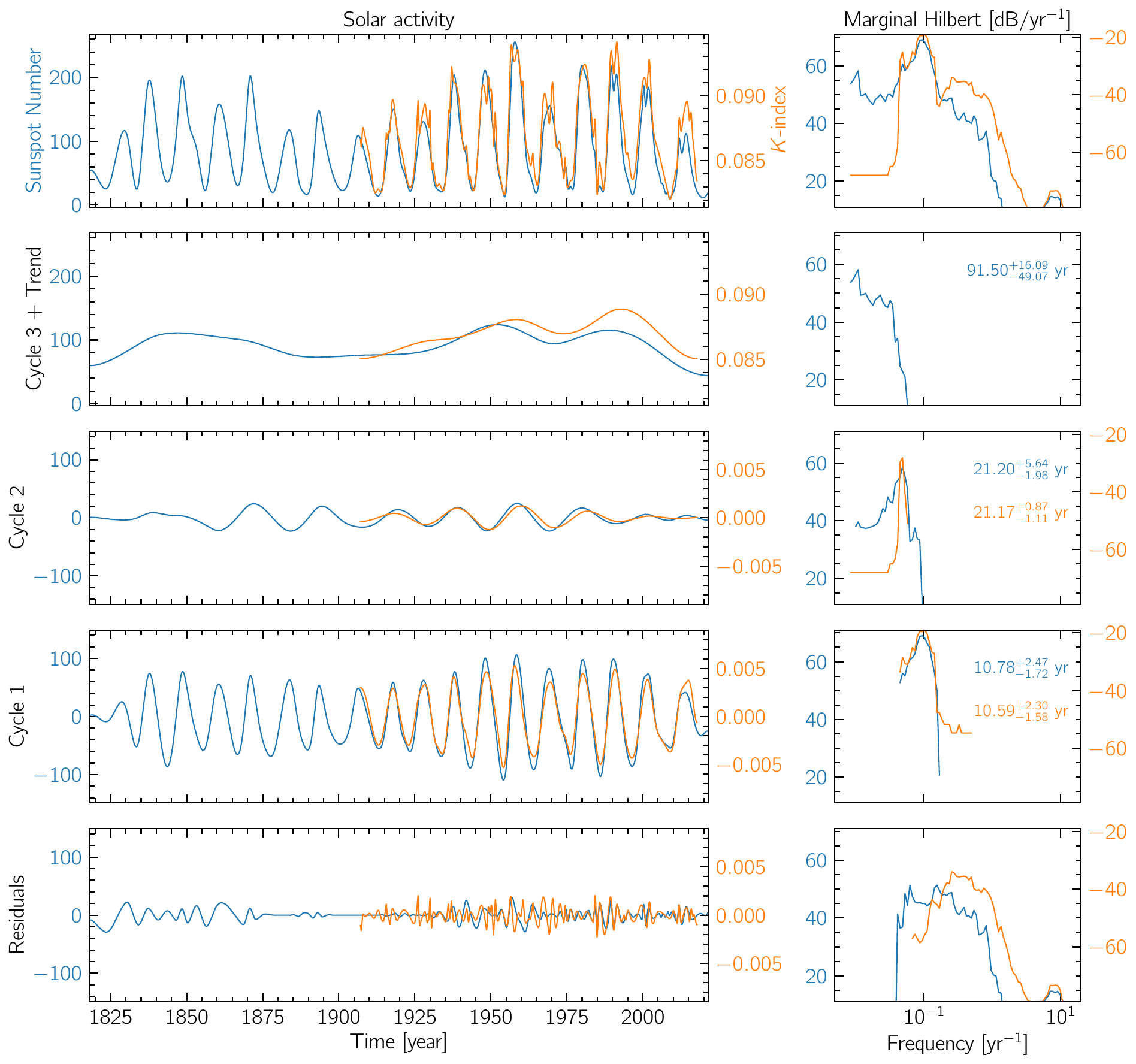}
\caption{Result of CEEMDAN for the Sun. The sunspot number is shown in blue and the $K$-index in orange. The top-left panel shows the preprocessed input data. The right-side panels show the Hilbert spectra corresponding to their respective time-domain components.}
\label{fig:sun}
\end{figure*}

To validate our methods, we tested the extent of our ability to recover multiple, distinct cycles on solar data using two different observational proxies, as detailed in Section \ref{sec:obs}: the daily total sunspot number series from 1818 to 2021, and the composite K-line emission index from 1907 to 2017 \citep{Bertello2016, Egeland2017}, hereafter referred to as the $K$-index.

Figure \ref{fig:sun} presents the result of our decomposition of solar activity data. The $K$-index is shown to correlate strongly with the sunspot number counts, and for both series our analysis identifies three main modes: the first one at $\approx$11 yr, another one at twice that period, and a longer $\approx$90-year cycle.
Although the $K$-index series is too short to resolve what appears to be the Gleissberg cycle, its trend component can be seen to correlate very well with the latter half of the final sunspot IMF, which indicates that this method is able to identify long-term trends and potential analogues of the solar Gleissberg cycle within the relatively short spectroscopic time series currently available for other stars.

It is reassuring to notice that the amplitude of the primary cycle seems to be modulated in a shape very similar to the Gleissberg component, which is to be expected if these components truly represent the underlying physical processes.
Also, the secondary cycle appears to capture the empirical Gnevyshev-Ohl rule, which dictates that odd solar cycles tend to be stronger than even ones \citep{Gnevyshev1948, Hathaway2015}.
The underlying cause of this observed behavior is thought to be related to the $\approx$22-year Hale magnetic cycle, but the connection is not yet entirely understood.

\subsection{61 Cygni A}

\begin{figure*}[ht!]
\plotone{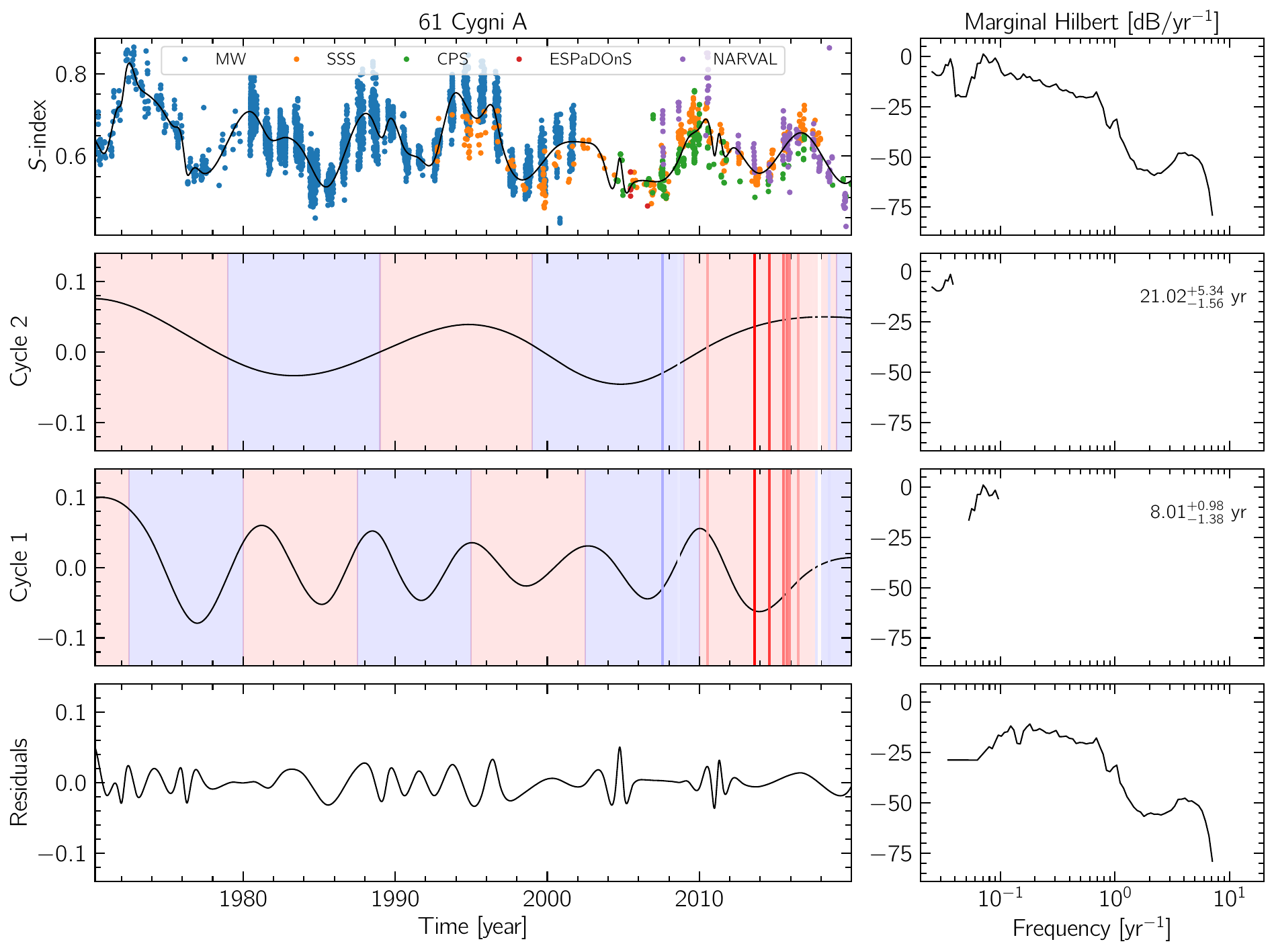}
\caption{Result of our analysis for 61 Cyg A. The colored vertical lines indicate the epochs for which its large scale magnetic geometry was reconstructed using ZDI \citep{BoroSaikia2016, BoroSaikia2018}. The background colors indicate predicted polarities for a hypothesis of a complete magnetic reversal taking  three times cycle 1 (shown in the Cycle 2 panel) or two times cycle 1 (as in the Sun, shown in the Cycle 1 panel).}
\label{fig:cyg}
\end{figure*}

We also applied our decomposition technique to the much shorter time series of 61 Cyg A $S$-index measurements.
The composition of observations from different instruments is shown in the top left panel of Figure \ref{fig:cyg} together with the smooth interpolation resulting from our preprocessing steps. The remaining panels of Figure \ref{fig:cyg} present the three best defined cyclic components found by the application of CEEMDAN and their corresponding Hilbert spectra.
Interestingly, besides the well-known $\approx$8-years long chromospheric activity cycle \citep[e.g.,][; cycle 1]{Baliunas1995}, a component with a period about three times as long is also well separated (cycle 2).
This component can be visually identified in the original data as a variation in activity maxima repeating itself in a pattern every three consecutive cycles. Such behavior is reminiscent of the solar Gnevyshvev-Ohl rule, except with a period $\approx$3 times rather than twice that of the activity cycle.

This star had its large-scale magnetic field detected via ZDI in six different epochs ranging from 2007 to 2015 \citep{BoroSaikia2016} and six more epochs from 2015 to 2018 \citep{BoroSaikia2018}. Its magnetic field polarity was found to flip twice between the first and last observation, indicating the presence of a Hale-like magnetic polarity cycle in this star.
Those polarity measurements are shown in Figure \ref{fig:cyg} as vertical lines, colored according approximately to the average field polarity and intensity at these epochs. An inversion can be seen to occur between the timeframes of two chromospheric activity minima, in agreement with the hypothesis of a $\approx15$-years long magnetic cycle. However, it is also important to note that a very short baseline was used to draw this inference, and the few datapoints we have cannot exclude the possibility that the Hale-like polarity cycle is actually linked to the observed three-cycle Gnevyshev-Ohl pattern, with a somewhat longer inversion period of $\approx11$ years.
The predictions of both these hypotheses are shown in Figure \ref{fig:cyg} as the colored red and blue backgrounds in the two cyclic components.

\section{Conclusions} \label{sec:conc}

We introduced a fully data-driven approach to separate different coexisting activity cycle components using the Hilbert-Huang transform that is immune to the non-stationary variations in cycle lengths and amplitudes known to exist.

This methodology was shown to be capable of not only recovering the mean periods of known solar cycles at different timescales, but also the time-domain characteristics of these components, such as amplitude variations, in agreement with expectations.

For the K-dwarf 61 Cygni A, our method confirms its Schwabe-like 8-year cycle and also presents evidence of a potential three-cycle long Gnevyshev-Ohl-like behavior, shown to be possibly tied to a polarity reversal. A confirmation of this result would be of some importance in the sense that (i) it strengthens the relationship between the solar Gnevyshev-Ohl rule and the Hale cycle, and (ii) it shows that we are able to detect the activity signature of magnetic polarity
changes in stellar spectroscopic calcium data.

\appendix

\section{On the Dependence of the Results on the Choice of Smoothing}

While describing our proposed pipeline and its main advantages, we highlighted the fact that it was a fully adaptive and data-driven method and that it doesn't rely on previous knowledge surrounding the dataset. We also claimed that, even though there apparently is need of a choice surrounding the smoothing procedures during the preprocessing steps, there was no strong effect on the results as long as the bin sizes were chosen within some reasonable interval.
This short appendix intends to justify such a claim.

\begin{figure}[ht]
    \centering
    \plotone{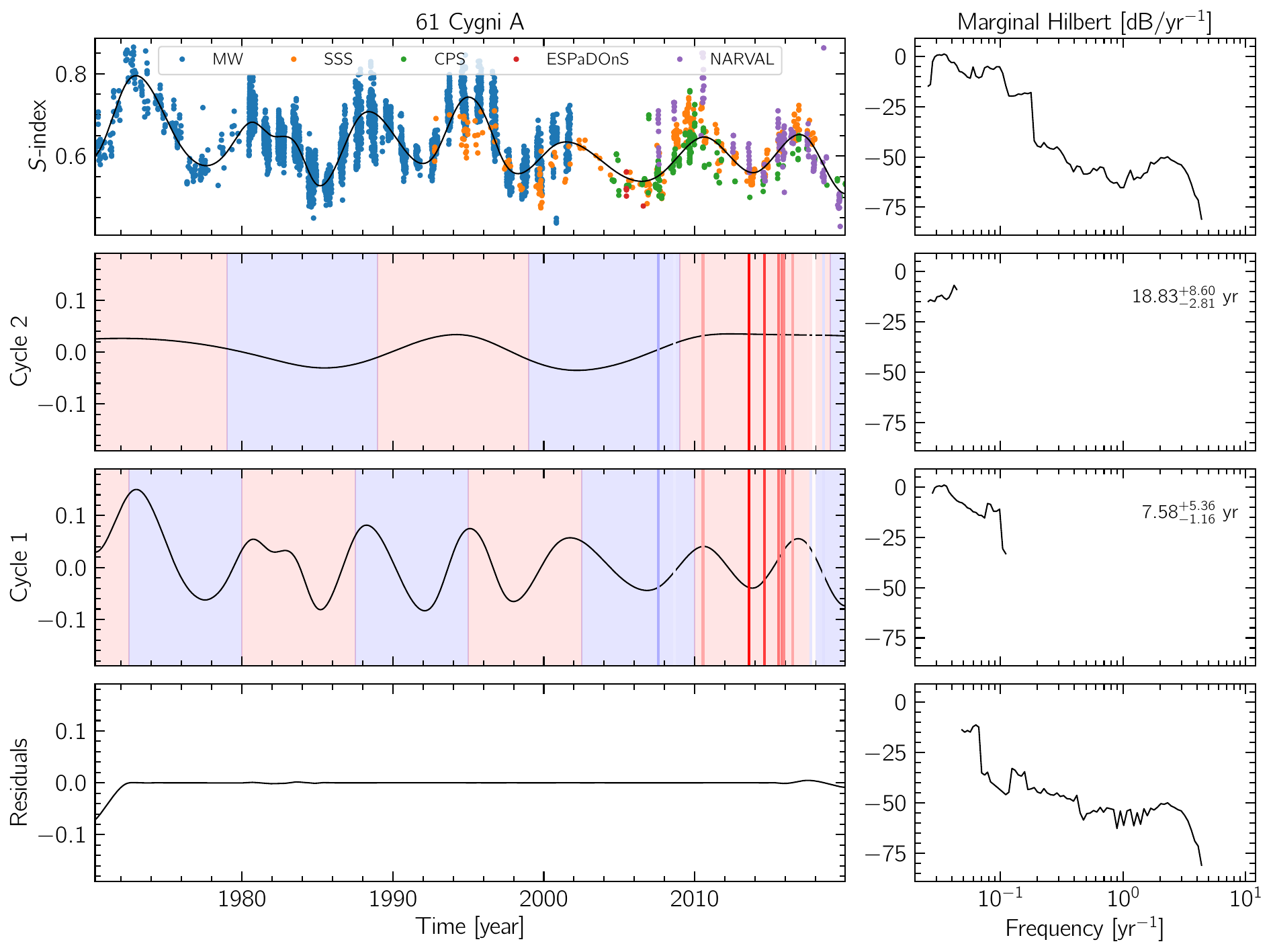}
    \caption{Same as Figure 2, however for a smoothed out version of the signal using 150-day bins during downsampling.}
    \label{fig:app}
\end{figure}

The main parameter that controls how much the input data is smoothed out is the size of the bins used for averaging. Figure \ref{fig:app} shows an example when the bins are almost twice as spaced out as what was used in the paper (see Figure \ref{fig:cyg}). The spikes get smoothed out and the high-frequency residuals found by the decomposition tend to disappear, but the overall structure of the two main cycles stays closely similar. The same qualitative conclusions can be taken out from this, thanks to the 
adaptive filtering nature of the method.

\begin{acknowledgments}
The HK\_Project\_v1995\_NSO data derives from the Mount Wilson Observatory HK Project, which was supported by both public and private funds through the Carnegie Observatories, the Mount Wilson Institute, and the Harvard-Smithsonian Center for Astrophysics starting in 1966 and continuing for over 36 years.  These data are the result of the dedicated work of O. Wilson, A. Vaughan, G. Preston, D. Duncan, S. Baliunas, and many others.  SHS is grateful for support from NASA Heliophysics LWS grant NNX16AB79G, NASA XRP grant 80NSSC21K0607, and NASA EPRV grant 80NSSC21K1037.
JDNJr acknowledges the financial support by CNPq funding through the grant PQ 310847/2019-2.
\end{acknowledgments}

\bibliography{sample631}{}
\bibliographystyle{aasjournal}

\end{document}